\documentclass[aps,pra,reprint,nofootinbib,superscriptaddress,twocolumn,showpacs,showkeys,longbibliography,amsmath,amssymb]{revtex4-2}
\usepackage{graphicx}
\usepackage{dcolumn}
\usepackage{bm}
\usepackage{braket}
\usepackage{subfigure}
\usepackage[colorlinks,bookmarks=false,citecolor=blue,linkcolor=red,urlcolor=blue]{hyperref}
\usepackage[english]{babel}
\usepackage{changes}
\usepackage{multirow}
\usepackage{booktabs}
\usepackage{float}
\usepackage{array}
\begin{document}
\title{Nonreciprocity-induced fractional nonlinear Thouless pumping}
\author{Yanqi Zheng}
\affiliation{Department of Physics, Zhejiang Normal University, Jinhua 321004, China}
\author{Kun Pu}
\affiliation{Department of Physics, Zhejiang Normal University, Jinhua 321004, China}
\author{Lingqing Ren}
\affiliation{Department of Physics, Zhejiang Normal University, Jinhua 321004, China}
\author{Chenxi Bai}
\affiliation{Department of Physics, Zhejiang Normal University, Jinhua 321004, China}
\author{Zhaoxin Liang}\email[Corresponding author:~] {zhxliang@zjnu.edu.cn}
\affiliation{Department of Physics, Zhejiang Normal University, Jinhua 321004, China}
\begin{abstract}
Recent interest has surged in eigenvalue nonlinearity-based topological transport governed by the equation of auxiliary eigenvalues $H\Psi=\omega S(\omega)\Psi$ [\href{https://doi.org/10.1103/PhysRevLett.132.126601}{ Isobe {\it et al.,} Phys. Rev. Lett. 132, 126601 (2024)}; \href{https://doi.org/10.1103/PhysRevA.111.042201}{ Bai and Liang, Phys. Rev. A 111,042201 (2025)}; \href{https://doi.org/10.1103/tgzz-1ngp}{Bai and Liang,  Phys. Rev. A 112, 052207 (2025)}] rather than the conventional Schr\"odinger equation $H\Psi=E\Psi$  in conservative settings, bit non-Hermitian generalizations remain uncharted. In this work, we are motivated to investigate the nonlinear Thouless pumping in a non-Hermitian and nonlinear Rice-Mele model. In particular, we uncover how non-Hermiticity parameters can induce fractional topological phases—even in the presence of quantized topological invariants as predicted by conventional linear approaches. Crucially, these fractional phases are naturally explained within the framework of the equation of auxiliary eigenvalues, directly linking nonlinear spectral characteristics to the bulk-boundary correspondence. Our findings reveal emergent phenomena arising from the interplay between nonlinearity and non-Hermiticity, providing key insights for the design of topological insulators and the controlled manipulation of quantum edge states in the real world.

DOI:\href{https://link.aps.org/doi/10.1103/w2sw-br7d}{10.1103/w2sw-br7d}
\end{abstract}
\maketitle

\section{INTRODUCTION\label{sec1}}

Thouless pumping~\cite{Thouless1983,NiuQ1984,Citro2023} in linear quantum systems serves as the paradigmatic example of quantized topological transport, where adiabatic cyclic modulation of parameters drives robust charge displacement governed by the TKNN relation~\cite{Thouless1982}. The physical mechanism~\cite{Di2010} behind the Thouless pumping demonstrates how topology manifests in periodically driven systems through quantized transport properties, establishing a foundational framework for understanding nontrivial band structures in quantum dynamics. Its experimental realizations encompass ultracold atomic gases~\cite{Lohse2016,Nakajima2021}, photonic systems~\cite{Kraus2012,Zilberberg2018}, and spin-based platforms~\cite{Ma2018}, demonstrating exceptional experimental versatility while preserving quantitative agreement with theoretical predictions. Next, nonlinearity, a key ingredient playing out in various distinct disciplines such as physics, biology, chemistry, economics, and social sciences, can induce nonlinear Thouless pumping~\cite{Jurgensen2021, Jurgensen2022, Fu2022a, Fu2022b, Mostaan2022, Tuloup2023, Hu2024, Lyu2024, Szameit2024, Cao2024a, Cao2024b, Cao2025,Burger1999, Denschlag2000, Strecker2002, Wu2003, Bleu2016, Watanabe2016, Lumer2013, Morimoto2016, Hadad2016, Zhou2017, Smirnova2020,Viebahn2024,Zhu2025}. In more details, nonlinearity induces quantized transport through soliton formation and spontaneous symmetry-breaking bifurcations, contrasting with linear Thouless pumping in electronic systems~\cite{Thouless1983,NiuQ1984,Citro2023} that necessitates complete occupation of an entire linear band. Meanwhile, quantized nonlinear Thouless pumping~\cite{Chen2025} can be understood by the Chern number of a Bloch band of the linear Hamiltonian, which constitutes a natural extension of the TKNN relation to nonlinear systems.  Additionally, fractional nonlinear Thouless pumping can occur for a soliton when the linear Hamiltonian is topologically nontrivial~\cite{Jurgensen2023} or trivial~\cite{Tao2025}, whose transport mechanisms unique to the existence of nonlinearity have gone beyond the framework of TKNN relation.  

Recently, experimental advances across distinct physical platforms, including photonics~\cite{StJean2017, Zeuner2015}, ultracold atoms~\cite{Xu2017, Li2020, Liang2023}, and acoustic systems~\cite{Zhu2018, Kawakami2016}, have enabled the controlled realization of non-Hermitian systems, characterized by engineered gain-loss mechanisms or nonreciprocal couplings. These systems exhibit unique topological phenomena absent in Hermitian settings, such as the non-Hermitian skin effect~\cite{Lee2016, Yao2018, Kunst2018, Yoshida2020a, Yokomizo2019, Borgnia2020, Okuma2020, Zhang2020} and exceptional points~\cite{Zhen2015, Shen2018, Yoshida2018, Zyuzin2018, Takata2018, Budich2019, Yoshida2019a, Yoshida2020b, Yoshida2019b, Okugawa2019, Zhou2019, Delplace2021, Mandal2021, Isobe2021, Isobe2023}. Owing to continuous advances in theoretical and experimental techniques~\cite{Malzard2015, Choi2010, Lee2014, Ashida2020, Gardas2016, Kawabata2023, Kawabata2019a, Li2019, Ren2022, Liang2022, Sun2023, Chen2021, Chen2022, Xiao2017, Wang2021, Ding2021, Gong2018, Edvardsson2019, Longhi2020, Pan2020, Li2021, Zhan2017, Hao2023, Kawabata2019b, Bergholtz2021, Poli2015, Miri2019, Ozdemir2019, Feng2022, Shu2022, Longhi2019}, non-Hermitian topological transport has been widely investigated~\cite{Kumar2025, Zhang2024, Ezawa2024, Kyriakou2021, Fedorova2020, Yuce2019, Hockendorf2020, Yang2025}. In these systems, the conventional bulk-boundary correspondence often needs to be extended using concepts like biorthogonal bases and generalized Brillouin zones~\cite{Yao2018, Guo2021, Ji2024, Yang2020} to correctly capture complex spectral features and the localization of skin modes.

While non-Hermiticity can induce linear topological phase transitions, existing research predominantly focuses on stabilizing linear responses. Recent breakthroughs in photonic platforms~\cite{Xia2021,Konotop2016} now enable exploration of non-Hermitian-nonlinear coupling regimes—a critical frontier due to the poorly understood interplay between these two dimensions. This integration introduces profound questions: How do nonlinearity and non-Hermiticity synergistically reshape topological transport mechanisms? Can their interaction engender novel bulk-boundary correspondence paradigms that transcend conventional quantum topology? Resolving these issues holds transformative potential for non-equilibrium quantum simulation, robust photonic devices, and fundamental understanding of open quantum matter.

\begin{figure}[!h] 
	\begin{centering} 
		\includegraphics[scale=0.25]{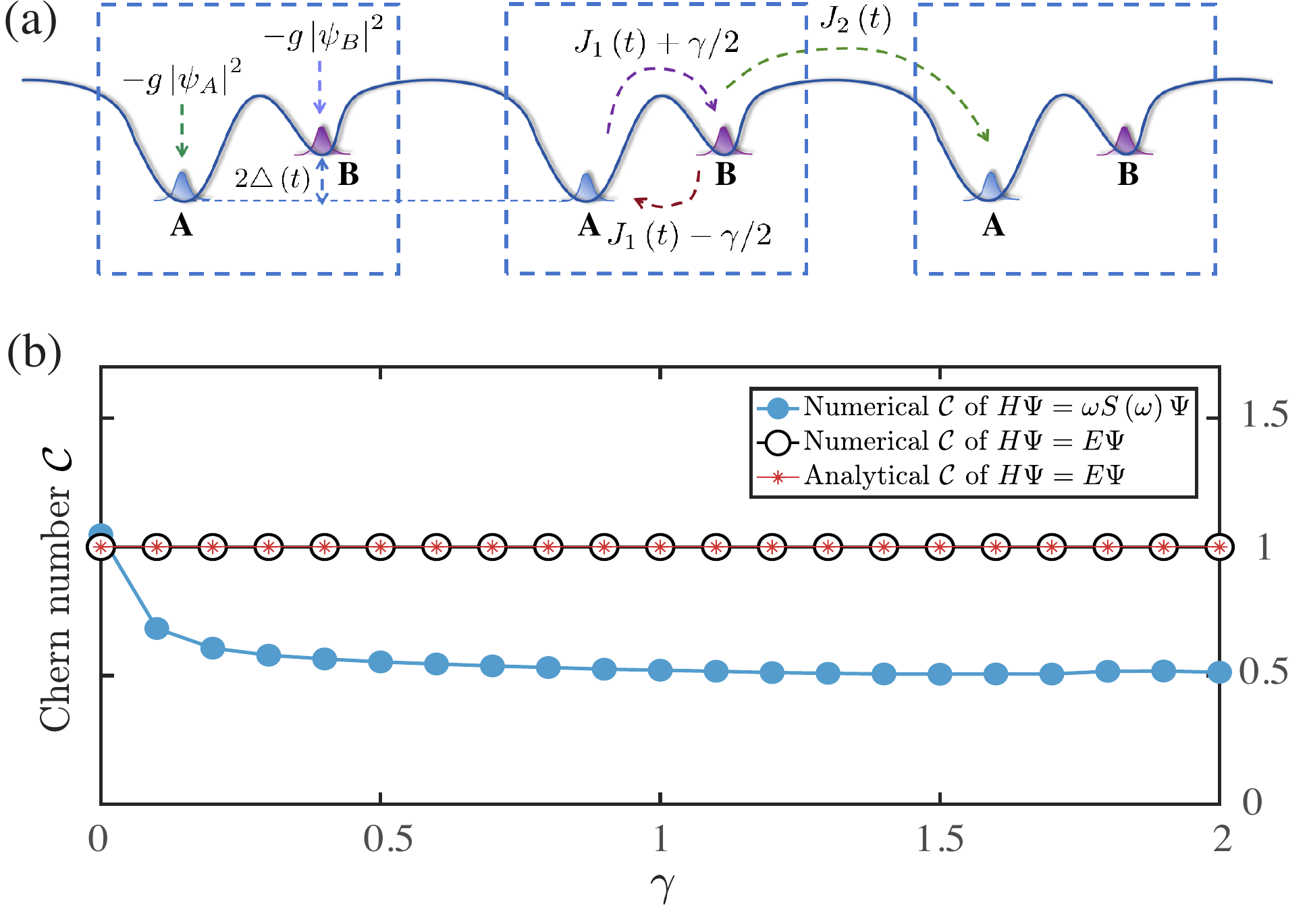} 
		\par\end{centering} 
	\caption{ (a) Schematic of the nonlinear non-Hermitian RM model. The dashed box indicates a unit cell consisting of sublattice A and B.
		(b) Chern number $\mathcal{C}$ as a function of $\gamma$ for $g=0$. The Chern numbers for the linear eigenvalue problem $H\Psi=E\Psi$, computed analytically and numerically, are shown by the red star and black open circle, respectively. The blue solid circles represent the Chern number for the nonlinear problem $H\Psi=\omega S(\omega)\Psi$, obtained numerically.
		The relevant parameters are given as: $J=1$, $\delta=0.5$, $\Delta=1$, $T=2000\pi$, and $\omega=\omega_d=10^{-3}$.
		\label{fig1}}
		 \vspace{-2mm}
\end{figure}

The second motivation of this work stems from the recent breakthrough in Refs.~\cite{Isobe2024,Bai2025a,Bai2025b}, which established a direct link between bulk-edge correspondence and eigenvalue nonlinearity in both linear~\cite{Isobe2024} and nonlinear~\cite{Bai2025a,Bai2025b} Hamiltonians through the auxiliary eigenvalue equation $H\Psi = \omega S(\omega)\Psi$. For example, wave propagation in dispersive media with frequency-dependent material response leads to a nonstandard eigenvalue problem, where the frequency (as the eigenvalue) appears in the operator~\cite{Deng2026,Ma2026}.
Conceptually going beyond previously conventional approaches~\cite{Thouless1983, NiuQ1984, Citro2023}, this formulation reveals how spectral properties of the bulk Hamiltonian  $H$ govern nonlinear eigenvalue problems, thereby bridging a unified description of bulk-edge physics through the lens of nonlinear eigenvalue problems in Hermitian physical systems. In particular, the auxiliary eigenvalue formulation of $H\Psi = \omega S(\omega)\Psi$ predict observable phenomena~\cite{Bai2025b} inaccessible to conventional linear bulk-edge paradigms~\cite{NiuQ1984}. Up to now, investigations of bulk-edge correspondence based on the framework of nonlinear eigenvalue problems via auxiliary eigenvalues have focused on Hermitian physical systems, however, their counterpart, the non-Hermitian scenario is largely unexplored.

To address the above-mentioned questions, we are motivated to employ a non-Hermitian nonlinear Rice-Mele (RM) model [Fig.~\ref{fig1}(a)] characterized by nonlinearity $g$ and non-Hermiticity $\gamma$, to systematically investigate topological transport in a non-Hermitian nonlinear environment.  We first separately calculate the relationship between the system's Chern number $\mathcal{C}$ and the non-Hermitian parameter $\gamma$ under linear ($H\Psi=E\Psi$) and nonlinear  [$H\Psi=\omega S\left(\omega\right)\Psi$] eigenvalue conditions [Fig.~\ref{fig1}(b)]. Surprisingly, we identify parameter regimes where the non-Hermiticity parameter $\gamma$ induces fractional topological phases, as explained within the auxiliary eigenvalue framework. This occurs even in the presence of quantized topological invariants—predicted by conventional linear approaches—highlighting a counterintuitive deviation from traditional topological paradigms. Subsequently, numerical simulations tracking soliton center-of-mass trajectories under adiabatic conditions reveal pumping dynamics where combined non-Hermitian and nonlinear parameters induce transitions from integer to fractional Chern phases [see Fig.~\ref{fig5}(a) below]. The latter phenomenon demonstrates how eigenvalue nonlinearities generate observable bulk-boundary correspondences absent in conventional linear frameworks, establishing a novel paradigm for topological transport in non-Hermitian and nonlinear systems. Our results predict that fractional non-Hermitian nonlinear Thouless pumping can be realized in photonic waveguides and cold-atom systems, opening different avenues for investigating topological states in non-equilibrium open systems.

This paper is structured as follows: In Sec.~\ref{sec2} A, we introduce the non-Hermitian nonlinear RM model. Sec.~\ref{sec2} B presents the calculation and analysis of the Chern number under both linear and nonlinear eigenvalue conditions. The derivation of the underlying nonlinear Schrödinger equation is detailed in Sec.~\ref{sec3} A, while Sec.~\ref{sec3} B employs the instantaneous soliton method to analyze the associated energy spectrum and phase diagram. Building upon this theoretical framework, Sec.~\ref{sec4} provides detailed numerical results of the soliton displacements and energy spectra, elucidating the distinct impacts of non-Hermiticity and nonlinearity. The bulk-boundary correspondence of the model and the regulatory roles of both non-Hermitian and nonlinear effects on topological edge states are elaborated in Sec.~\ref{sec5}. Finally, Sec.~\ref{sec6} summarizes the main findings and discusses their implications for future studies.

\section{NONLINEAR NON-HERMITIAN RICE-MELE MODEL\label{sec2}}
\subsection{Model system}
In this work, we are interested in a one-dimensional interacting bosonic chain containing N dimerized units [see Fig.~\ref{fig1}(a)], where each unit consists of two distinct lattice sites labeled A and B. Within mean-field approximation, the system's static and dynamic properties are governed by a non-Hermitian generalization of the nonlinear RM Hamiltonian $H_{\text{RM}}$~\cite{Tuloup2023,Bai2025a,Bai2025b}, augmented with additional non-Hermitian terms $H_{\text{NH}}$ proposed in Ref.~\cite{Yang2025}. The total Hamiltonian $H_{\text{NRM}}=H_{\text{RM}}+H_{\text{NH}}$ of our model system then reads
\begin{eqnarray}
	H_{\text{RM}} &=&\sum_{n=1}^{N-1}\left(J_{1}\Psi_{n,A}^{*}\Psi_{n,B}+J_{2}\Psi_{n,B}^{*}\Psi_{n+1,A}+\text{H.c.}\right)\nonumber \\
	&-&\Delta\cos\left(\omega_{\text{d}} t\right)\sum_{n=1}^{N}\left(\left|\Psi_{n,A}\right|^{2}-\left|\Psi_{n,B}\right|^{2}\right)\nonumber \\
	&-&\frac{g}{2}\sum_{n=1}^{N}\left(\left|\Psi_{n,A}\right|^{4}+\left|\Psi_{n,B}\right|^{4}\right), \label{RM}\\
	H_{\text{NH}}
	&=&\sum_{n=1}^{N}\frac{\gamma}{2}\left(\Psi_{n,A}^{*}\Psi_{n,B}-\Psi_{n,B}^{*}\Psi_{n,A}\right)\label{NH}.
\end{eqnarray}

In Eqs.~(\ref{RM}) and (\ref{NH}), each unit cell consists of two sublattices denoted as A and B, where $\Psi_{n,A}$ and $\Psi_{n,B}$ correspond to the wavefunction on the A and B sublattices within the $n$-th unit cell, respectively. The time-dependent intra-cell and inter-cell hopping strengths are given by $J_{1}=-J-\delta\sin\left(\omega_{d} t\right)$ and $J_{2}=-J+\delta\sin\left(\omega_{d} t\right)$. The parameter $J$ sets the uniform hopping amplitude, $\delta$ controls the modulation amplitude, and $\omega_{d}$ is the modulation frequency. To ensure adiabatic evolution, we require $\omega_{d} \ll J, \delta$.  The term $\Delta\cos\left(\omega_{d} t\right)$ represents a time-varying step-like on-site potential with magnitude $\Delta$. Nonlinear interactions are incorporated via a Kerr-type term of strength $g$~\cite{Tuloup2023,Bai2025a,Bai2025b}.  The coefficient $\gamma$ in Eq. (\ref{NH}) quantifies the non-reciprocal character of the hopping process. The emphasis of this work is on investigating how the non-Hermitian nature captured by Hamiltonian (\ref{NH}) can affect the topological properties of a nonlinear RM model of Hamiltonian (\ref{RM}). 

Before exploring non-Hermitian effects [labeled by $\gamma$ in Eq. (\ref{NH})] on the nonlinear RM model, we would like to establish the basic properties~\cite{Tuloup2023,Bai2025a,Bai2025b} of the nonlinear RM model by setting $\gamma=0$. Compared with linear RM model~\cite{Rice1982}, the introduction of the nonlinearity of $g$ in Eq. (\ref{RM}) has brought rich dynamics into topological phases, typical of nonlinear integer~\cite{Jurgensen2021, Jurgensen2022, Fu2022a, Fu2022b, Mostaan2022, Tuloup2023, Hu2024, Lyu2024, Szameit2024, Cao2024a, Cao2024b, Cao2025,Burger1999, Denschlag2000, Strecker2002, Wu2003, Bleu2016, Watanabe2016, Lumer2013, Morimoto2016, Hadad2016, Zhou2017, Smirnova2020} and fractional Thouless pumping~\cite{Jurgensen2023,Tao2025}. In more detail, nonlinearity acts to quantize transport via soliton formation and spontaneous symmetry-breaking bifurcations. In the exploration of the  physics arising from the interplay of nonlinearity and topology, the Thouless pumping of such solitons has been experimentally observed\cite{Jurgensen2021,Viebahn2024,Zhu2025}, and theoretically studied in both the weak nonlinear regime and the strong nonlinear regime\cite{Fu2022a, Fu2022b, Mostaan2022, Tuloup2023,Bai2025a,Bai2025b}.

Furthermore, incorporating additional non-Hermitian terms described by Hamiltonian (\ref{NH}) into the nonlinear RM model of Hamiltonian (\ref{RM}) induces non-equilibrium dynamics, manifesting two key features: (1) Topological invariants require biorthogonal-basis calculations, potentially altering system topology, and (2) non-Hermitian skin effects localize eigenstates at boundaries. These points immediately suggest two novel implications of nonlinear Thouless pumping: (1) Emergent bulk-boundary correspondences in non-Hermitian nonlinear regimes, and (2) skin-characterized solitons under specific parameter conditions. These predictions align with recent experiments realizing nonlinear skin solitons~\cite{Wang2025} and Hopf-bifurcation-induced criticality in non-Hermitian skin effects~\cite{Kawabata2025}.

\subsection{Non-Hermitian Chern number}
The topological properties of the nonlinear non-Hermitian RM model described by Eqs. (\ref{RM}) and (\ref{NH})  are governed by four key parameters: nearest-neighbor coupling strengths $J_1$ and $J_2$, the nonreciprocal parameter $\gamma$, and nonlinear interaction strength $g$. This work reveals that the synergistic interplay between non-Hermitian skin effects and nonlinear band engineering unveils novel bulk-edge correspondences, enabling the exploration of exotic phenomena beyond conventional linear Hermitian frameworks. Specifically, we demonstrate that $\gamma$-induced directional amplification and $g$-mediated interaction-induced gap openings govern phase transitions through a competitive mechanism, while establishing topologically protected channels for solitonic excitations. These results establish a paradigm for engineering non-Hermitian topological phases via coupled nonlinear-dispersion mechanisms, holding promise for photonic and quantum simulator platforms with reconfigurable boundary states.


As the first step, we derive the topological invariants for the linearized non-Hermitian RM model with $g=0$ in Eq.~(\ref{RM}). For periodic boundary conditions, the momentum-space wavefunctions for each sublattice follow from discrete Fourier transformation as $\left|\Psi_{A(B)}\left(k,t\right)\right\rangle =1/\sqrt{N}\sum_{j=1}^{N}e^{ikj}\left|\Psi_{j,A(B)}\left(t\right)\right\rangle $ with $k=2\pi m/N$ and $m=0,1,\dots,N-1$.
The total wavefunction is expressed as a two-component spinor $\left|\Psi\left(k,t\right)\right\rangle =\left(\left|\Psi_{A}\left(k,t\right)\right\rangle ,\left|\Psi_{B}\left(k,t\right)\right\rangle \right)^{T}$, and then the momentum-space form of  Hamiltonian (\ref{RM}) and (\ref{NH}) can be written in the form of $H=\sum_k\left\langle \Psi\left(k,t\right)\right|H\left(k,t\right)\left|\Psi\left(k,t\right)\right\rangle $. Here, the single-particle Hamiltonian $H\left(k,t\right)$ can be written as
\begin{equation}
	H\left(k,t\right)=d_{x}\sigma_{x}+(d_{y}+i\frac{\gamma}{2})\sigma_{y}+d_{z}\sigma_{z},\label{OneH}
\end{equation}
with $d_{x}=J_{1}+J_{2}\cos k$, $d_{y}=J_{2}\sin k$, $d_{z}=-\Delta\cos\omega t$ and $\sigma_{x,y,z}$ representing the Pauli matrices. The topological properties of our model system can be characterized by the Chern number 
\begin{equation}
	\mathcal{C}=\frac{1}{2\pi}\int_{0}^{2\pi/\omega_{\text{d}}}dt\int_{-\pi}^{\pi}dk\Omega\left(k,t\right),\label{chern}
\end{equation}
where $\Omega\left(k,t\right)$ is the Berry curvature. In non-Hermitian systems, we adopt the definition of $\Omega\left(k,t\right)$~\cite{Lee2016, Yao2018, Kunst2018, Yoshida2020a, Yokomizo2019, Borgnia2020, Okuma2020, Zhang2020} in the biorthogonal basis as detailed in Appendix \ref{AppendixA}
\begin{eqnarray}
	\Omega
	 =i\left\langle \partial_{t}u_{\text{L}}\right|\left.\partial_{k}u_{\text{R}}\right\rangle
	 -i\left\langle \partial_{k}u_{\text{L}}\right|\left.\partial_{t}u_{\text{R}}\right\rangle.\label{Berry curvature}
\end{eqnarray}
Here, $\left|u_{\text{R}}\left(k,t\right)\right\rangle$ and $\left|u_{\text{L}}\left(k,t\right)\right\rangle$ are the eigenstates of $H\left(k,t\right)$ and $H^{\dagger}\left(k,t\right)$ in Eq. (\ref{OneH}), respectively. In general, as long as the energy gap of Hamiltonian (\ref{OneH}) remains open during the adiabatic evolution, the Chern number $\mathcal{C}$ defined in Eq. (\ref{chern}) can ensure the quantization of the charge transfer in each pumping cycle. 

Next, we need to obtain the biorthogonal eigenstates of $|u_{\text{L(R)}}\rangle$ in Eq. (\ref{Berry curvature}) in order to calculate the Chern number in Eq. (\ref{chern}). As pointed out in Refs.~\cite{Isobe2024, Cheng2024, Bai2025a, Bai2025b},  two kinds of scenarios of calculating $|u_{\text{L(R)}}\rangle$  can be adopted 

(1) In the first scenario, one can obtain the biorthogonal eigenstates of $|u_{\text{L(R)}}\rangle$ based on the linear Hamiltonian $H\Psi=E\Psi$ in the traditional quantum mechanics.  Based on the definitions of Chern number $\mathcal{C}$ in Eq. (\ref{chern}), we proceed to analytically calculate the Chern numbers for different values of $\gamma$, as shown by the red stars in Fig.~\ref{fig1}(b). We can see from the Fig. .~\ref{fig1}(b) that the system always remains in a topologically nontrivial state with $\mathcal{C}=1$. To verify the analytical results, we also employ a numerical method based on the Wilson loop to calculate the Chern number~\cite{Bai2025a,Bai2025b}, the results of which are represented by the black open circles in Fig.~\ref{fig1}(b). The consistency between the Chern numbers obtained with the two methods indicates that the Wilson-loop method~\cite{Alexandradinata2014} is not only reliable but also more convenient in practical calculations and can accurately reflect the topological properties of the system.

(2) The second scenario is referred to as the nonlinear eigenvalue equation defined as follows~\cite{Isobe2024,Bai2025a,Bai2025b}:
\begin{equation}
	H\left(k,t\right)\Psi=\omega S\left(\omega \right)\Psi,\label{nh=ws}
\end{equation}
with $S\left(\omega\right)$ being the overlap matrix and $\omega$ representing the eigenvalue's nonlinearity. Equation~(\ref{nh=ws}) can describe wave propagation in dispersive media with frequency-dependent material response, yielding a nonstandard eigenvalue problem in which the frequency~(eigenvalue) appears in the eigenvalue-equation operator~\cite{Ma2026,Deng2026}.
Then we define the auxiliary matrix $P\left(\omega,k\right)=H\left(k,t\right)-\omega S\left(\omega\right)$. At this point, the nonlinear equation~(\ref{nh=ws}) is transformed into the auxiliary equation $P\left(\omega,k\right)\Psi=\lambda\Psi$. Here, $S\left(\omega\right)=I-M_{S}\sigma_{z}$, and $M_{S}\left(\omega\right)=M_{1}\tanh\left(\omega t\right)$, where $M_{S}\left(\omega\right)$ reflects the dependence on nonlinearity. At this stage, the frequency $\omega$ becomes a parameter, while the eigenvalues $\lambda$ of the auxiliary equation form the energy bands.  Only when $\lambda=0$ do the solutions of the auxiliary equation coincide with those of Eq.~(\ref{nh=ws}), and the corresponding $\omega$ represents the physical eigenfrequency of the system.
From a physical standpoint, therefore, we need to focus on the case $\lambda=0$. By constructing an auxiliary linear eigenvalue problem, we can analyze and predict the topological edge states of the physical nonlinear eigenvalue problem, thereby establishing a bulk-edge correspondence tailored to nonlinear systems. The Chern numbers calculated via this second method for different $\gamma$ values are shown as blue circles in Fig.~\ref{fig1}(b).

\begin{figure}[!t]
	\begin{centering} 
		\includegraphics[scale=0.52]{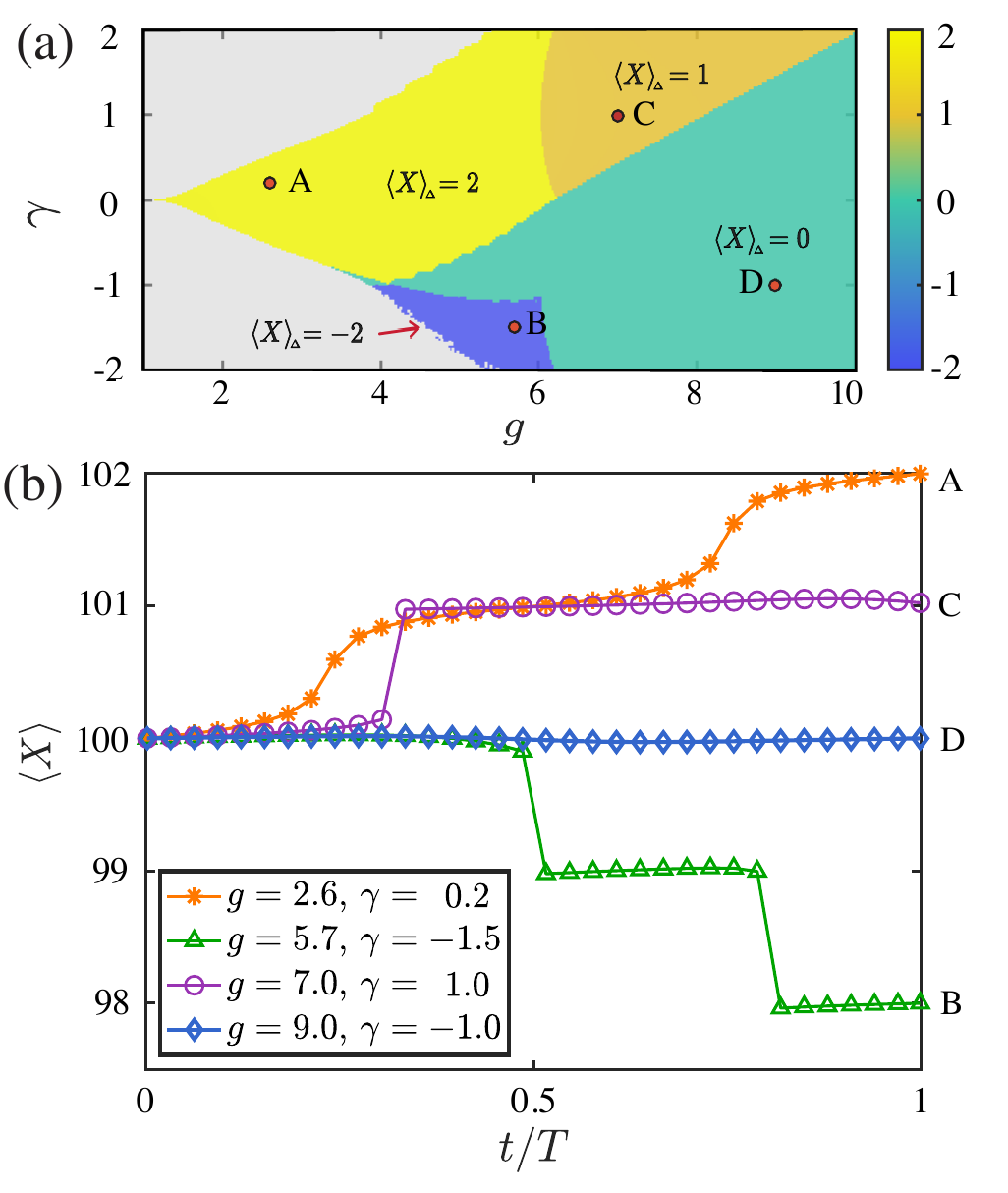} 
		\par\end{centering} 
	\caption{ (a) The phase diagram of the soliton centroid transport variation calculated under the scanned parameters $\left(g,\gamma\right)$. (b) The centroid transport conditions corresponding to the four points marked in Fig. 2(a). The relevant parameters are given as: $A\left(2.6,0.2\right)$, $B\left(5.7,-1.5\right)$,  $C\left(7.0,1.0\right)$, and $D\left(9.0,-1.0\right)$. Fixed parameters are: $J=1$, $\delta=0.5$, $\Delta=1$, $T=2000\pi$, and $\omega=10^{-3}$.
		\label{fig2}
       \vspace{0mm}
	}
\end{figure}

Finally, the results for the Chern number in Eq. (\ref{chern}) calculated in the two scenarios above are displayed in Fig.~\ref{fig1}(b). In more detail, the red stars represent the analytical Chern numbers based on linear eigenvalues, while the black open circles and blue solid circles represent the numerical Chern numbers corresponding to linear eigenvalues and nonlinear eigenvalues, respectively. It is worth noting that the calculations of both linear and nonlinear eigenvalues are performed under the condition of $g=0$. As can be seen from Fig.~\ref{fig1}(b), the Chern number remains at $\mathcal{C}=1$ in the linear case, while under the nonlinear eigenvalue problem, the Chern number exhibits fractional behavior and gradually stabilizes at $\mathcal{C}=\frac{1}{2}$. This behavior can be understood as the geometric phase of the effective Bloch bands being reconstructed under the combined action of strong nonlinearity and non-Hermitian coupling, leading to the coupling and excitation of multiple topological bands and resulting in fractional Chern numbers. Based on the above results, when non-Hermitian nonlinearity is introduced [blue circles in Fig.~\ref{fig1}(b)], solitons~\cite{Bai2025b} are expected to exhibit fractional pumping phenomena. In contrast, such fractional pumping behavior is completely absent in the linear case [red stars and black circles in Fig.~\ref{fig1}(b)].

We remark that, to the our best knowledge,  the fractional Thouless pumping of solitons was first explored in Ref. \cite{Jurgensen2023}: At weak nonlinearity, soliton motion retains quantization characterized by the Chern number—this arises because solitons remain tightly associated with single-band Wannier centers throughout the pumping cycle. In contrast, under stronger nonlinearity, solitons couple to multi-band Wannier functions instead, which leads to the emergence of fractional quantization in their transport properties. In what follows, we elaborate on the distinctions in the physical origins governing the quantized fractional Thouless pumping of solitons between our study and Ref. \cite{Jurgensen2023}.  The fractional pumping phenomenon reported in this work stems from the introduction of non-Hermitian $\gamma$ and is elucidated within the auxiliary eigenvalue framework, rather than the conventional multiband Wannier tracking mechanism driven by strong interactions. As explicitly demonstrated in Fig. \ref{fig2}(a), we have carefully verified that merely adjusting the nonlinearity—without including nonreciprocity ($\gamma=0$)—fails to induce fractional Thouless pumping. This conclusively shows that the predicted fractional pumping cannot be accounted for by the physical picture of solitons coupling to multi-band Wannier functions via strong interactions. Fractional Thouless pumping emerges exclusively when the parameter $\gamma$ become nonzero and is tuned into the regime marked by $\mathcal{C} = \frac{1}{2}$ in Fig. \ref{fig2}(a), as illustrated in Fig. \ref{fig2}(b). We therefore conclude that the fractional pumping phenomenon studied here originates from the inclusion of nonreciprocity ($\gamma\neq0$), with its mechanism explained by the auxiliary eigenvalue framework.

\section{Topological phase diagram of non-Hermitian and nonlinear RM model \label{sec3}}
\subsection{Nonlinear Schrödinger equation}
In Sec.~\ref{sec2}, we mentioned that for the non-Hermitian nonlinear Hamiltonian with $g\neq 0$, the nonlinear term can form solitons within the lattice. Therefore, in the scenario illustrated by Fig.~\ref{fig1}(a) (blue circles), we predict that solitons will exhibit fractional pumping behavior. Our strategy for demonstrating the physical manifestations of the fractional Chern number within the nonlinear framework [Fig. \ref{fig1}(b)] is inspired by the concept of nonlinear Thouless pumping~\cite{Jurgensen2021, Jurgensen2022, Fu2022a, Fu2022b, Mostaan2022, Tuloup2023, Citro2023, Hu2024, Lyu2024, Szameit2024, Cao2024a, Cao2024b, Cao2025}. Specifically, for the non‑Hermitian nonlinear RM Hamiltonian (1) with $g\neq 0$, the nonlinearity quantizes transport through soliton formation and nonuniform band occupation. Hence, solitons are expected to undergo topological pumping in the scenario described. Moreover, the physical mechanism behind the soliton pumping can be understood from the Chern number computed from the corresponding linear Hamiltonian with $g=0$. As far as we know, calculating the topological invariant for nonlinear systems—using the nonlinear eigenstates of the nonlinear equation—remains an outstanding challenge and lies beyond the scope of the present work. Hence, we will leave the calculation of the Chern number from the Hamiltonian $H_{\text{NRM}}$ with $g\neq 0$ in Eqs. (\ref{RM}) and (\ref{NH}) for future study.

Specifically, we expect that the soliton supported by Hamiltonian (\ref{RM}) and (\ref{NH}) will undergo fractional pumping in the scenario depicted in Fig. \ref{fig1}(b).

To investigate the dynamical behavior of solitons governed by Hamiltonians (\ref{RM}) and (\ref{NH}), we derive the corresponding equations of motion using $i\partial\Psi_{j}/\partial t=\delta H/\delta\Psi_{j}^{*}$~\cite{Bai2025a,Bai2025b}. Specifically, the system evolution of $\Psi_{j}$ is governed by the discrete nonlinear non-Hermitian Schrödinger equations as follows,
\begin{eqnarray}
	i\frac{\partial\Psi_{j,A}}{\partial t} 
	& = & -\left(J+\delta\sin\omega_d t-\frac{\gamma}{2}\right)\Psi_{j,B} \nonumber\\
	& & \quad -\left(J-\delta\sin\omega_d t+\frac{\gamma}{2}\right)\Psi_{j-1,B} \nonumber\\
	& & \quad -\left(\Delta\cos\omega_d t+g|\Psi_{j,A}|^{2}\right)\Psi_{j,A}, \label{non-H schrodinger eqa1}\\
	i\frac{\partial\Psi_{j,B}}{\partial t} 
	& = & -\left(J+\delta\sin\omega_d t-\frac{\gamma}{2}\right)\Psi_{j,A} \nonumber\\
	& & \quad -\left(J-\delta\sin\omega_d t+\frac{\gamma}{2}\right)\Psi_{j-1,A} \nonumber\\
	& & \quad +\left(\Delta\cos\omega_d t-g|\Psi_{j,B}|^{2}\right)\Psi_{j,B}, \label{non-H schrodinger eqa2}
\end{eqnarray}
where $j=1,2\cdots N-1,N$ labels the lattice sites. Equations.~(\ref{non-H schrodinger eqa1}) and (\ref{non-H schrodinger eqa2}) define a discrete nonlinear non-Hermitian Schrödinger system, which is fundamentally equivalent to the lattice Gross-Pitaevskii equation describing Bose systems with nonlinear interactions and non-Hermitian characteristics. This framework effectively models the nonequilibrium dynamics of ultracold atomic gases in optical lattices and optical pulse propagation in photonic crystals with engineered gain-loss profiles~\cite{Viebahn2024,Zhu2025}. 


\subsection{Soliton displacement phase diagram}
In this section, we plan to focus on the two phenomena predicted by the Chern number in Fig. \ref{fig1}(b), i.e. the quantized and fractional nonlinear Thouless pumping by numerically solving Eqs.~(\ref{non-H schrodinger eqa1}) and ~(\ref{non-H schrodinger eqa2}) with the fixed parameters $J=1$, $\delta=0.5$, and $\Delta=1$ within the parameter regimes of $1\leqslant g\leqslant 10$ and $-2\leqslant\gamma\leqslant 2$. The system employs open boundary conditions ($\Psi_{1}=\Psi_{2N}=0$), and the lattice size is $N=100$ cells (a total of 200 lattice sites). To verify the reliability of the numerical results, we have ensured that all simulation results are independent of system size through finite-size analysis. All physical quantities in the text have been nondimensionalized. At the initial moment, a soliton state is placed at lattice site $n=100$. The construction of this soliton state employs the iterative self-consistent algorithm proposed in the literature, with the trial wave function taken as $\Psi_{0}=\text{sech}\left[\left|x-100\right|/5\right]$. The soliton wave function at each moment is then calculated using the equations. For specific implementation details, see Appendix~\ref{AppendixB}.

Based on the numerical methods~\cite{Bai2025a,Bai2025b}, we can study the pumping dynamics of the soliton by monitoring the evolution of its centroid position over time, where the centroid position is defined as $\left\langle X\right\rangle =\sum_{j}j\left|\Psi_{j}\right|^{2}$. We remark that there are two methods to calculate  $\langle X\rangle=\sum_j j|\Psi_j|^2$: dynamical evolution and instantaneous solution of solitons. For the former, we evolve an initial soliton to time $t$ via fourth-order Runge-Kutta integration of the wavefunction's time evolution. For the latter, we find the soliton at $t$ iteratively to steady state. Under adiabatic conditions, both converge as expected, mutually verifying results. Under nonadiabatic conditions from interaction-induced self-intersecting bands, causing discrepancies between instantaneous and dynamical simulations, the dynamical method reflects experimental reality, while the instantaneous one diagnoses adiabaticity breakdown. Details of the numerical methods are in Ref.~\cite{Bai2025a,Bai2025b}.

In Fig.~\ref{fig2}, we present the topological phase diagrams by calculating the soliton's displacement of $\left\langle X\right\rangle _{\Delta}$ through numerically solving  Eqs. (\ref{non-H schrodinger eqa1}) and (\ref{non-H schrodinger eqa2}) with different values of  
the nonlinearity $g$ and the non-Hermitian parameter $\gamma$. As demonstrated in Fig. \ref{fig2}(a), nonlinear Thouless pumping exhibits four distinct transport regimes, color-coded according to $\left\langle X\right\rangle_{\Delta}=\pm 2$, $1$ and $0$. These regimes correspond to three kinds of fundamentally different dynamical responses under periodic parameter modulation.  The case of quantized nonlinear Thouless pumping characterized $\left\langle X\right\rangle_{\Delta}=\pm 2$ in Fig. \ref{fig2}(a)  means that the soliton is transported by two sites per cycle  and corresponds to Chern number $\mathcal{C}=\pm 1$ [see the curve with black circles in Fig. \ref{fig1}(b)], satisfying with the TKNN relation. In contrast, the case of fractional nonlinear Thouless pumping characterized $\left\langle X\right\rangle_{\Delta}=1$ in Fig. \ref{fig2}(a)  means that the soliton is transported by one site per cycle  and corresponds to Chern number $\mathcal{C}=\pm \frac{1}{2}$ [see blue curve in Fig. \ref{fig1}(b)], breaking the TKNN relation through the interplay between $g$ and $\gamma$. Note that the gray area corresponds to the non-Hermitian skin effect of the soliton's transport. In more detail, we select four points in the four distinct regions above, labeled $A$, $B$, $C$, and $D$, and monitor how the soliton's displacements change over time within one pumping cycle in Fig.~\ref{fig2}(b). It is evident that the soliton's displacements of $\left\langle X\right\rangle_{\Delta}$ indeed are transported to $\left\langle X\right\rangle_{\Delta}=\pm 2$, $1$ and $0$ as expected.


\section{Non-Hermitian-induced fractional Thouless pumping in non-Hermitian and nonlinear RM model\label{sec4}}

\begin{widetext}
	
	\begin{figure}[t] 
		\begin{centering} 
			\includegraphics[scale=0.38]{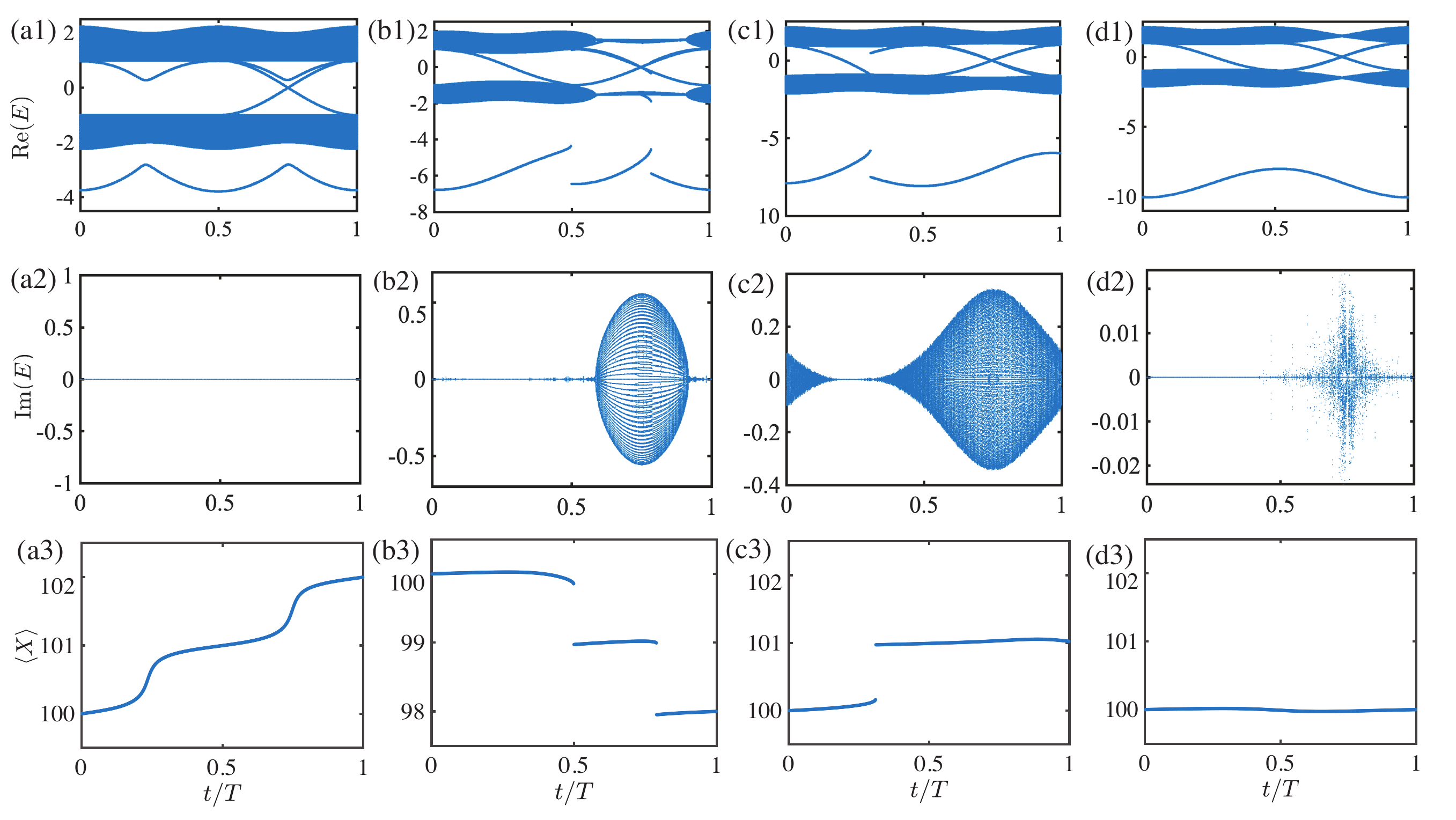} 
			\par\end{centering} 
			\caption{Energy spectrum and soliton dynamics for the nonlinear non-Hermitian RM Hamiltonian. (a1)–(d1): Real energy spectrum structure, (a2)–(d2) Complex energy spectrum structure. (a3)–(d3) Time evolution of the soliton centroid position expectation value $\langle X \rangle$ over one period.  The interaction strength $g$ and non-Hermitian strength $\gamma$ are set as follows: (a) $g=2.6$, $\gamma=0.2$; (b) $g=5.7$, $\gamma=-1.5$; (c) $g=7.0$, $\gamma=1.0$; and (d) $g=9.0$, $\gamma=-1.0$. These cases correspond respectively to the points labeled $A$, $B$, $C$, and $D$ in Fig. 2(a). Parameter values are fixed at $J=1$, $\delta=0.5$, $\Delta=1$, $T=2000\pi$, and $\omega=10^{-3}$.
			\label{fig3}
            \vspace{2mm}
		}
	\end{figure}
	
\end{widetext}

In Sec.~\ref{sec3}, we introduced the discrete non-Hermitian nonlinear Schrödinger equations (\ref{non-H schrodinger eqa1}) and ~(\ref{non-H schrodinger eqa2}), and drew the phase diagram of soliton displacement varying with parameters based on these two equations. In this section, we will more carefully analyze the effects of non-Hermiticity on the energy spectrum and soliton transport behavior of the nonlinear RM model and explore the role that non-Hermiticity plays. 

First, we analyze the real and complex energy spectra alongside soliton centroid trajectories for parameters at points A–D in Fig.~\ref{fig2}(a), demonstrating four distinct dynamical regimes. In the weakly nonlinear regime [small $g$, Figs.~\ref{fig3}(a1)-3(a3)], the energy spectrum remains entirely real. Remarkably, introducing a finite non-Hermitian parameter $\gamma$ preserves the soliton displacement of two over one pumping cycle, identical to the Hermitian case. This invariance indicates that topological transport properties remain robust against weak non-Hermiticity in this nonlinear regime.


The quantized transport of solitons observed in Fig.~\ref{fig3}(a3) directly arises from the Chern number of the non-interacting band structure ($g=0$). In the weakly nonlinear regime, solitons localize in the lowest energy band [$\mathcal{C}=1$], see Fig.~\ref{fig1}(b)), maintaining quantized displacement through bulk-boundary correspondence. This manifests as a quantized displacement of two lattice sites per pumping cycle, demonstrating the robustness of topological transport against non-Hermitian perturbations. At strong nonlinearities [Fig.~\ref{fig3}(d3)], band splitting induces Rabi oscillations between split bands, revealing the breakdown of topological protection. Since the total Chern number of the two bands sums to zero, the net displacement of the soliton vanishes, and the transport process is completely suppressed. In the moderate-nonlinearity regime, however, the position jumps observed in the soliton trajectory [see Fig.~\ref{fig3}(d3)] arise from the ring structure of the bands, which disrupts the adiabatic evolution path. Introducing non-Hermiticity at this point induces significant non-Hermitian effects: Under the parameters in Fig.~\ref{fig3}(d), while the absolute value of the soliton transport distance remains unchanged, its direction reverses, and the energy spectrum acquires an imaginary component. This phenomenon underscores the critical role of non-Hermiticity in shaping the system's dynamical behavior—specifically, the complexification of the energy spectrum and the reversal of the pumping direction are directly tied to the modulation of the Bloch band's geometric phase by the non-Hermitian parameter.


Fig.~\ref{fig3}(c3) demonstrates non-Hermiticity-induced fractional soliton pumping, where the soliton exhibits half-integer displacement per pumping cycle under the parameters shown. This behavior emerges from altered bulk-boundary correspondence that develops when non-Hermiticity modifies the system's topology, creating a distinct transport regime absent in the linear regime. The fractional pumping arises from the interplay between non-Hermitian skin effects and modified band structures, as predicted by the non-Bloch Chern number calculations in Fig.~\ref{fig1}(b).


\begin{figure}[t] 
	\begin{centering} 
		\includegraphics[scale=0.37]{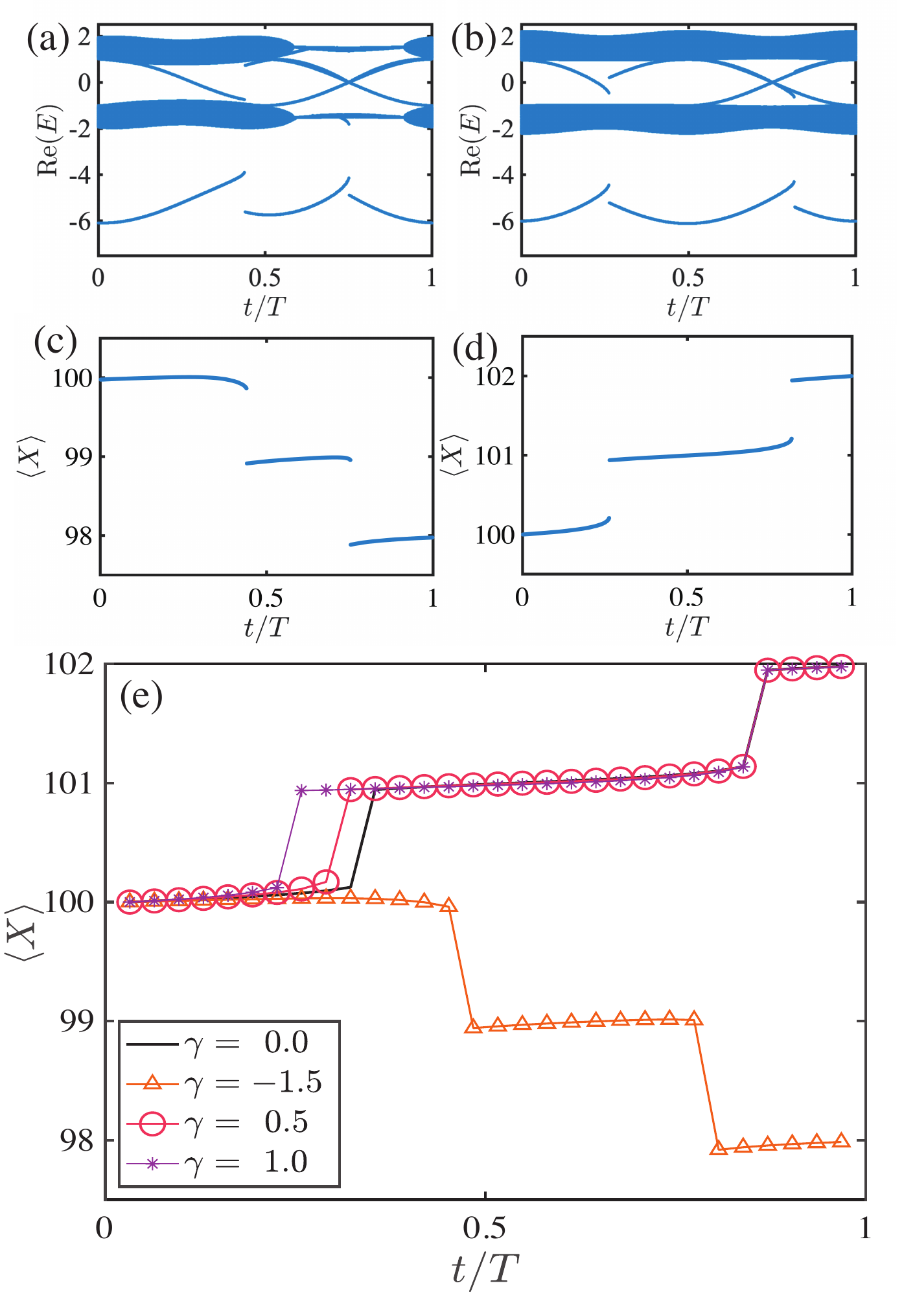} 
		\par\end{centering} 
	\caption{ Non-Hermitian control of topological transport. (a) and (b) Real part of the non-Hermitian nonlinear eigenvalue spectrum; (c)–(e) Quantized soliton dynamics. Fixed parameters are $g=5$, $J=1$, $\delta=0.5$, $\Delta=1$, $T=2000\pi$, and $\omega=10^{-3}$.
		\label{fig4}
	}
\end{figure}\begin{figure}[t] 
	\begin{centering} 
		\includegraphics[scale=0.37]{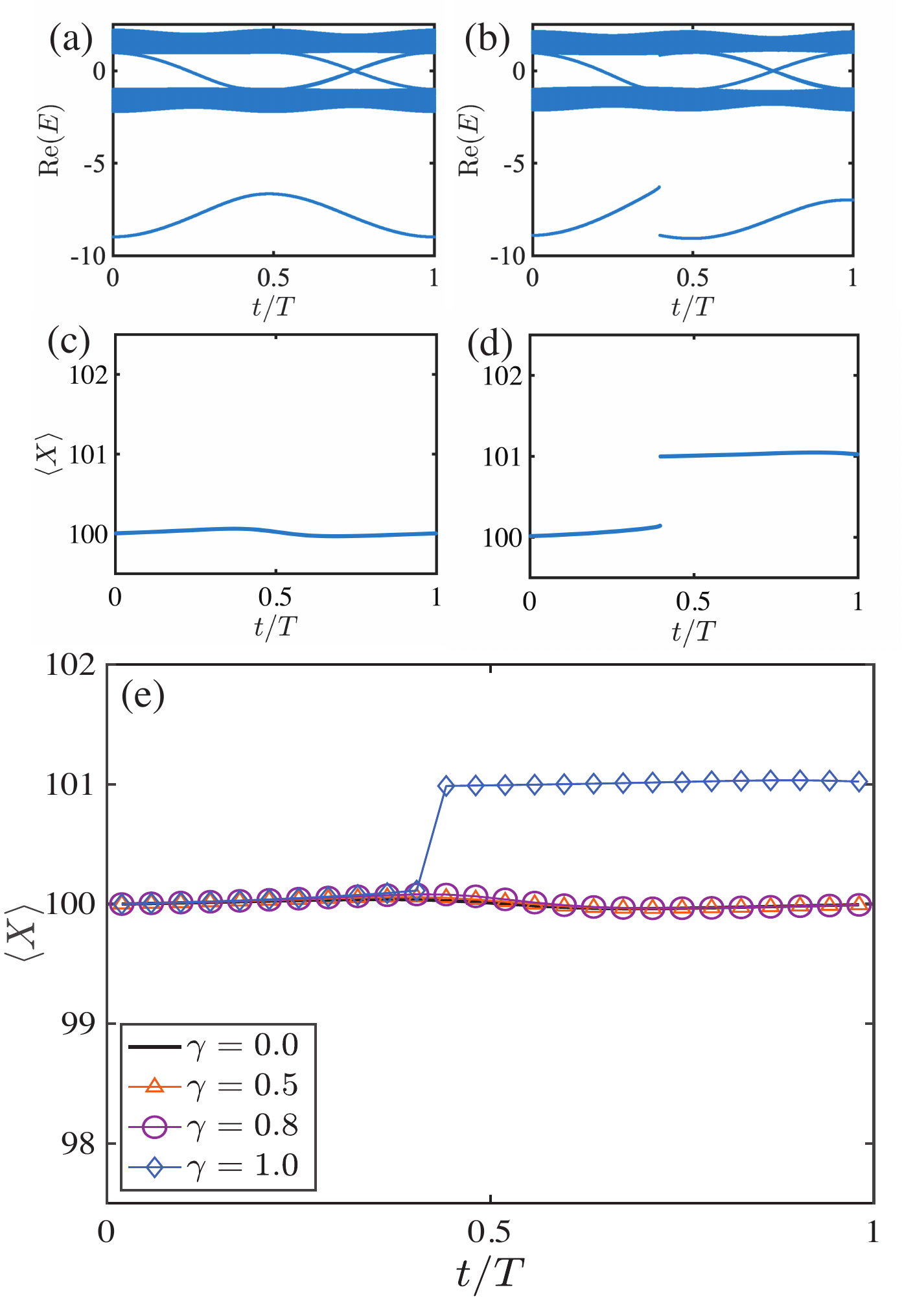} 
		\par\end{centering} 
	\caption{ Non-Hermitian modulation of topological transport: (a) and (b) Real part of the non-Hermitian nonlinear eigenvalue spectrum; (c,d,e) Quantized soliton dynamics. Fixed parameters are: $g=8$, $J=1$, $\delta=0.5$, $\Delta=1$, $T=2000\pi$, and $\omega=10^{-3}$.
		\label{fig5}
	}
\end{figure}

Next, following the observation of the predicted fractional transport phenomenon in Fig. \ref{fig1}(b) and \ref{fig2}(a), we investigate the effect of the non-Hermitian parameter $\gamma$ in Eq.~(\ref{NH}) on the quantized transport of solitons while keeping the moderate nonlinearity fixed at $g=5$. Our systematic analysis in Fig. \ref{fig4} reveals three key regimes: (1) In the Hermitian limit of $\gamma=0$, the system exhibits the typical nonlinear RM model behavior as shown by the black solid curve in Fig. \ref{fig4}(e), with position jumps during soliton pumping arising from band self-intersections induced by nonlinearity. (2) In the moderate regime of $\gamma=0$, as the value of $\gamma$ increases [see the curves with circles and triangles in Fig. \ref{fig4}(e)], the pumping behavior remains stable, with integer-quantized displacement of two lattice sites per cycle. This indicates that moderate non-Hermiticity does not alter the system's underlying topological properties, consistent with the robustness of bulk-boundary correspondence in non-Hermitian topological systems. (3) In the regime where $\gamma<0$,  when $\gamma$ decreases below a critical threshold, the pumping direction reverses while the absolute value of the displacement remains $2$, indicating that non-Hermiticity mainly affects the transport direction without changing its quantized characteristics.


Finally, we present results for the regime where bulk-boundary correspondence undergoes qualitative changes, studied at strong nonlinearity ($g=8$). At $\gamma=0$ in the Hermitian limit, the system exhibits significant soliton localization, and topological pumping is completely suppressed. Introducing a finite non-Hermiticity ($\gamma\neq 0$) reactivates topological transport, even in this strongly nonlinear regime. For subthreshold values of  $\gamma$ (below a critical threshold), the system retains behavior analogous to the Hermitian case, with negligible changes in soliton dynamics. However, when $\gamma$ exceeds this threshold, the system enters a fractional quantum pumping regime, characterized by half-integer displacement of solitons per pumping cycle. The transport trajectories obtained via the instantaneous soliton method (Fig.~\ref{fig5}(d)) clearly visualize this fractional pumping process, providing direct evidence that non-Hermitian terms can reconstruct the system's effective topological structure in the presence of strong nonlinearity. This reconstruction enables the emergence of fractional transport phenomena, which are absent in the Hermitian limit and highlight the profound impact of non-Hermiticity on topological pumping in nonlinear systems.


\section{NON-HERMITIAN NONLINEAR BULK-BOUNDARY CORRESPONDENCE\label{sec5}}

In this section, we systematically investigate the energy spectrum structure and soliton pumping images of the system under different parameters in the non-Hermitian nonlinear RM model. This section will further elaborate on the corresponding physical mechanism behind it—the bulk-edge correspondence.

Based on Eqs.~(\ref{non-H schrodinger eqa1}) and ~(\ref{non-H schrodinger eqa2}), we drew the phase diagram shown in Fig.~\ref{fig2} by plotting the soliton displacement results obtained by scanning the parameters $\left(g,\gamma\right)$. 
In the weakly and moderately nonlinear regions, changing the non-Hermitian parameter $\gamma$ does not affect the quantized value of the pumping.The integer quantization behavior can be predicted by the Chern number in the linear limit [corresponding to $\mathcal{C}=1$ indicated by the black circles in Fig~\ref{fig1}(b)]. In the strongly nonlinear region, however, adjusting $\gamma$ can induce fractional quantum transport of the soliton. In this case, the pumping quantity is no longer an integer and is predicted by the Chern number calculated from the nonlinear eigenvalues [corresponding to the $\mathcal{C}=\frac{1}{2}$ indicated by the blue circles in Fig.~\ref{fig1}(b)].

It is particularly noteworthy that the emergence of fractional transport strongly depends on the combined effect of nonlinearity and non-Hermiticity. As shown in Figs.~\ref{fig4} (e) and \ref{fig5} (e), in the Hermitian case ($\gamma=0$), adjusting the nonlinearity strength $g$ alone does not induce fractional pumping. Correspondingly, under weakly nonlinear conditions, even in the presence of non-Hermiticity, the system still exhibits integer quantum transport as shown in Fig.~\ref{fig4} (e). This indicates that only when strong nonlinearity and non-Hermiticity coexist can the band structure and geometric properties of the wave functions of the system be changed sufficiently significantly to support topologically nontrivial fractional pumping behavior as displayed by the diamond curve in Fig.~\ref{fig5} (e).

\section{CONCLUSION AND OUTLOOK\label{sec6}}

In summary, we investigate the energy spectrum characteristics and soliton dynamics of the nonlinear non-Hermitian RM model, highlighting non-Hermitian effects. By analyzing soliton dynamics via time-dependent evolution equations under combined nonlinear and non-Hermitian conditions, we observed fractionalized topological transport emerging when strong nonlinearity coexists with finite non-Hermitian strength. This phenomenon defies conventional linear and Hermitian frameworks but can be quantitatively predicted by non-Hermitian Chern numbers derived from auxiliary eigenvalues. Our results establish a generalized bulk-edge correspondence for nonlinear non-Hermitian systems, revealing how nonlinear interactions synergize with non-Hermitian effects to induce exotic topological transport. These findings advance the understanding of non-equilibrium topological phases and provide a theoretical foundation for controlling topological transport in photonic waveguides and ultracold atomic systems through combined nonlinearity and non-Hermiticity.

We remark that this work together with Refs. \cite{Isobe2024,Bai2025a,Bai2025b} establishes a comprehensive description of nonlinear eigenvalue-driven topological transport through the auxiliary eigenvalue equation
 $H\Psi=\omega S(\omega)\Psi$ in both conservative and nonconservative settings. This approach elucidates the synergistic mechanisms through which interactions, topology, and dissipation collectively enable novel nonequilibrium paradigms for topological phenomena. The derived bulk-edge correspondence demonstrates how nonlinear spectral characteristics govern boundary state emergence, providing a predictive framework absent in conventional linear theories. These findings advance the understanding of nonequilibrium quantum matter and offer a systematic methodology for engineering topological transport in photonic and atomic systems through combined nonlinearity and non-Hermiticity. In this work, we focused on fractional Thouless pumping in the nonreciprocal nonlinear RM model. Exploring similar phenomena in other nonlinear topological models---characterized by the interplay of nonlinearity and nonreciprocal hoppingg—is a promising direction. We will pursue such phenomena in other physical models in future work.
 

\section*{ACKNOWLEDGEMENTS}

We thank Ying Hu, Yapeng Zhang, Xuzhen Cao, Shujie
Cheng, and Biao Wu for stimulating discussions and useful
help. We sincerely acknowledge Yucan Yan for his dedicated
efforts during the critical inception phase of this project. This work was supported by the National Natural Science Foundation of China under Grant No. 12574301, the Zhejiang Provincial Natural Science Foundation of China under Grant No. LZ25A040004 and the Key Project of the National Natural Science Foundation of China Joint Funds under Grant No. U25A20197.
\appendix

\section{Chern number}\label{AppendixA}

In this appendix, we provide the definitions of the Berry curvature in Eq.~(\ref{Berry curvature}) according to the biorthogonal basis of $\left|u\left(k\right)\right\rangle_{R}$ and $\left|u\left(k\right)\right\rangle_{L}$, which read
\begin{eqnarray}
	\Omega_{kt}^{\alpha\beta}&=&i\left\langle \partial_t u^{\alpha}\left(k,t\right)\right|\left.\partial_k u^{\beta}\left(k,t\right)\right\rangle \nonumber \\
	&-&i\left\langle \partial_k u^{\alpha}\left(k,t\right)\right|\left.\partial_t u^{\beta}\left(k,t\right)\right\rangle ,
\end{eqnarray}
with the normalization condition $\langle u_{k}^{\alpha}|u_{k}^{\beta}\rangle =1$ and $\alpha,\beta\in\left\{ L,R\right\} $. When numerically calculating the Chern number, we employed an equivalent variant of Eq.~(\ref{chern}).  Previous studies ~\cite{Shen2018, Zhang2024} demonstrated that, despite having different definitions of these Berry curvatures $\Omega_{kt}^{\alpha\beta}$, they yield the same Chern number when integrated over the momentum-time manifold. For example, we have
\begin{equation}
	\mathcal{C}=\int_{0}^{\frac{2\pi}{\omega_{\text{d}}}}dt\int_{-\pi}^{\pi}dk\Omega_{kt}^{RR}=\int_{0}^{\frac{2\pi}{\omega_{\text{d}}}}dt\int_{-\pi}^{\pi}dk\Omega_{kt}^{LR}.
\end{equation}
Using $\Omega_{kt}^{RR}$ for the calculation will providemany conveniences.

\section{Numerical method} \label{AppendixB}

To find the steady-state solutions of Eqs.~(\ref{non-H schrodinger eqa1}) and ~(\ref{non-H schrodinger eqa2}) at any given time, we employ a self-consistent iterative method. For a given non-Hermitian nonlinear Hamiltonian $H$, the iterative steps from state $\left|\Psi_{n}\left(t\right)\right\rangle $ to state $\left|\Psi_{n+1}\left(t\right)\right\rangle $ at a specific time $t$ are as follows:

First, based on $\left|\Psi_{n}\left(t\right)\right\rangle $, one can obtain the corresponding instantaneous Hamiltonian of $H_{n}=H\left(\left|\Psi_{n}\right\rangle ,t\right)$ at the given time $t$.

Next, one can solve for the eigenstates $\left|\psi_{i}\right\rangle $ of $H_{n}$ with $i=q,\dots,2N$.

Finally, one can select the eigenstate $\left|\psi_{i}\right\rangle $ with the largest overlap with the previous eigenstate $\left|\Psi_{n}\right\rangle $ as the new state, i. e. for all $i$, $\left\langle \Psi_{n}\left(t\right)\right|\left.\psi_{i0}\right\rangle \geq\left\langle \Psi_{n}\left(t\right)\right|\left.\psi_{i}\right\rangle $. . Then one should repeat the iterative process until the overlap between the new state and the previous state reaches $1-\left\langle \Psi_{n+1}\left(t\right)\right|\left.\Psi_{n}\left(t\right)\right\rangle <10^{-10}$. The method is highly sensitive to the choice of the initial state, so the selection of the initial state must be done with caution. We verified the robustness of the nonlinear Thouless pumping against different initial-function choices, e.g., $\exp(-| x-100|)$ and $\text{sech}(| x-100|)$.

We remark that when the non-Hermitian strength exceeds a critical threshold, the non-Hermitian skin effect dominates, resulting in only the boundary-localized state with the largest overlap persisting, i. e. the skin effect causes the pumped soliton to accumulate at the system boundary. This phenomenon explains the emergence of skin solitons at the boundary and the origin of the gray region in Fig.~\ref{fig2}. The non-Hermitian and nonlinear effects exhibit competitive interplay: under skin-soliton-producing conditions, increasing the nonlinearity strength suppresses boundary localization, enabling the observation of solitons at the system center. 

\bibliography{zyqref}
\end{document}